# Modelling dynamics of samples exposed to free-electron-laser radiation with Boltzmann equations.

B. Ziaja [*,†] [1], A. R. B. de Castro [**], E. Weckert [*] and T. Möller [‡] [2]

[*] *HASYLAB at DESY, Notkestr. 85, D-22603 Hamburg, Germany*

[†] *Department of Theoretical Physics, Institute of Nuclear Physics, Radzikowskiego 152, 31-342 Cracow, Poland*

[**] *IFGW UNICAMP, CEP 13083-970, Campinas, SP Brasil;*

*LNLS, P. O. Box 6192, CEP 13084-971, Campinas, SP Brasil*

[‡] *Technische Universitaet Berlin, Institut fuer Atomare Physik und Fachdidaktik, 10623 Berlin, Hardenbergstrasse 36, Germany*

**Abstract:** We apply Boltzmann equations for modelling the radiation damage in samples irradiated by photons from free electron laser (FEL). We test this method in a study case of a spherically symmetric xenon cluster irradiated with VUV FEL photons. The results obtained demonstrate the potential of the Boltzmann method for describing the complex and non-equilibrium dynamics of samples exposed to FEL radiation.

## 1 Introduction

The emerging free-electron-lasers (FELs) promise tremendous progress in studying the structure of matter with soft and hard X-rays. The transversely fully coherent radiation from the

[1]Author to whom correspondence should be addressed

[2]Electronic mails: ziaja@mail.desy.de, arbcastro@lnls.br, edgar.weckert@desy.de, thomas.moeller@physik.tu-berlin.de

FEL will be delivered in flashes of ultrashort duration, emitted at peak brilliances more than $10^8$ higher than those available from the present sources of synchrotron radiation [1, 2]. These unique properties of FELs enable probing dynamic states of matter, transitions and reactions happening within tens of femtoseconds, with wide-ranging implications to the solid state physics, material sciences, and to femtochemistry. The FEL beam, if focussed onto a very small spot, is also an excellent tool to generate and probe extreme states of matter. The X-ray FEL is expected to open new horizons in the structural studies of biological systems, especially in the studies of non-repetitive samples, like viruses, living cells etc. Rapid progress of radiation damage in these samples prevents an accurate determination of their structure in standard diffraction experiments. However, the recent studies of the progress of the damage formation [3–5] indicate that the radiation tolerance might be extended at ultrafast imaging with high radiation dose as that expected with the presently developed X-ray FELs (LCLS, DESY).

For this and other applications of FELs, we have to understand in detail how the intense radiation of short wavelengths, emitted in short pulses, interacts with matter. This interaction has considerably different properties than the interaction of matter with photons emitted from optical lasers. At different energies of the FEL photons different processes are contributing to the radiation damage. At VUV photon energies the inverse bremsstrahlung process is believed to deliver most of the energy needed for the efficient ionization of the irradiated system [6]. This ionization eventually leads to the Coulomb explosion of the sample. This occurs, when the positively-charged sample is so strongly ionized that the repulsive electrostatic forces between ions move the atomic cores of the sample apart. Photons of even shorter wavelengths may excite electrons from inner shells of atoms, creating core holes [7]. Photoemissions, core-hole creations and subsequent Auger emissions of secondary electrons contribute to the radiation damage that then affects not only the sample but also the optical elements of the FEL beamline.

Radiation damage by photons from the VUV FEL is now under intense investigation. First experimental studies on the interaction of the VUV photon beams with atoms, molecules and clusters have been already performed at DESY [8]. The large number of VUV photons absorbed per atom observed in these experiments could not be explained using the well-established standard calculations for photon absorption [6, 8]. This indicated that the ionization of the sample irradiated by energetic photons progresses in a different way than that observed in the optical energy range. These surprising results stimulated intense theoretical effort. Several interesting models have been proposed [6, 9–11] which could explain various aspects of the increased photoabsorption and ionization dynamics observed in the experiments. On the other hand, there are still some controversies, e. g. regarding the role of the inverse bremsstrahlung mechanism and the inner ionization process. A model: (i) computationally efficient also for large spatially non-uniform samples and (ii) able to test the influence of specific interactions on the complex dynamics of electrons and ions, could be useful in evaluating the contributions of different mechanisms to the overall ionization dynamics.

Here we propose such efficient method of describing the evolution of irradiated samples which applies also to large systems. This first-principles Boltzmann method is based on the statistical description of the charge dynamics in terms of statistical quantities: electron and ion densities in phase space, $\rho^{(e,i)}(\mathbf{r}, \mathbf{v}, t)$. These densities are functions of the spatial and velocity coordinates, $\mathbf{r}$ and $\mathbf{v}$, and are measured at some time, $t$. The quantity, $\rho^{(e,i)}(\mathbf{r}, \mathbf{v}, t) d^3r\, d^3v$, estimates a number of particles (electrons or ions) in an infinitesimal volume element of phase space, $dV = d^3r\, d^3v$, which is located at the spatial point, $\mathbf{r}$, and at the velocity, $\mathbf{v}$. Charge densities are evolved from their initial configuration at $t = 0$, using semiclassical Boltzmann equations.

The Boltzmann method is a promising alternative to the first-principles Monte Carlo (MC) or Molecular Dynamics (MD) methods which are commonly used [3, 4]. The Monte Carlo method may be viewed as an approximate stochastic method of solving transport equations. During the simulation the MC code has to solve the separate equations of motion for each particle in the sample, following the trajectories of all particles and their interactions with other particles or external fields. If any scatterings occur, the scattering probabilities are estimated with quantum mechanical cross sections. Coordinates and velocities of particles are updated at each time-step. Simulations of single events are repeated many times. Estimates of physical observables are obtained by averaging their values obtained from single events over the total number of events. Therefore these estimates are biased with statistical errors.

Monte Carlo algorithms have a transparent structure and usually do not require an application of any complex numerical methods. This is a great advantage of this method. However, these algorithms become computationally inefficient when the number of particles, $N$, is large. The code efficiency is even worse, if the long-range interactions between particles (e.g. Coulomb electrostatic forces) have to be included into the simulation. High computational costs which scale with the number of particles restrict the applicability of the Monte Carlo method to the samples of small or moderate sizes.

In contrast, the efficiency of the simulation algorithm with the Boltzmann equations does not change directly with the number of particles in the sample, as the algorithm operates on smooth density functions. Therefore the efficiency (and accuracy) of these algorithms depend only on the phase-space shape of the sample which is reflected by the number of grid points used in the simulation. Therefore this statistical approach can also work fine for large samples, where the MD/MC methods are inefficient. Moreover, the full spatio-temporal characteristics of the electron and ion dynamics can be easily obtained with the transport method. As the charge densities are directly evolved with Boltzmann equations, the averaged observables, $O$, of interest can then be calculated with by their convolution with the charge densities obtained, $\langle O(t)\rangle = \int O(\mathbf{r},\mathbf{v})\,\rho(\mathbf{r},\mathbf{v},t)d^3r\,d^3v$. These results are not biased with statistical errors.

The applicability of Boltzmann equations is, however, limited to the systems which fulfill the assumptions of molecular chaos and two-body collisions. These assumptions are usually justified by a presence of short range forces [12, 13]. The single particle density function obtained with these equations does not contain any information on three-body and higher correlations. If the higher order correlations are important, a more fundamental Liouville equation for the N-particle density function should be applied. The Liouville equation reduces to the collisionless Vlasov equation [12] in case of an uncorrelated system. Fokker-Planck equation [12] can be derived as a limiting form of the Liouville equation for long-range (Coulomb) forces. It was proven in Ref. [12] that a correct description of many body electron-ion interactions in plasma obtained with the dedicated Fokker-Planck equations can be also obtained with the Boltzmann two-body collision term, assuming the Debye cutoff in the Rutherford scattering cross section. This simplification does not apply to the electron-electron interactions where the interacting charged particles have identical masses.

Another disadvantage of the statistical approach is its numerical complexity. Boltzmann equations are complicated sets of nonlinear integro-differential equations where partial derivatives appear in both spatial and in velocity coordinates, $\partial/\partial\mathbf{r}$, and $\partial/\partial\mathbf{v}$. Any approach to solve these equations requires a careful treatment of the initial and boundary conditions. It requires also maintaining the stability conditions which are necessary for the convergence of the algorithm. Advanced numerical methods have then to be applied.

In what follows we will show the potential of the Boltzmann method for studies of the radiation damage in samples irradiated by FEL photons. At the present state we do not aim to obtain any detailed and quantitative predictions which could be later compared to the existing experimental data. Actually, we are interested in proposing a new theoretical approach for a comprehensive and efficient description of the radiation damage in irradiated samples, almost independently of the sample size. A construction of a realistic model including all relevant physical processes is planned at later stages.

First we will write general Boltzmann equations for samples irradiated with VUV photons. These equations will include basic physical processes contributing at those photon energies. We will then solve these equations in a simplified (study) case of a spherically symmetric xenon cluster. Due to the symmetries of the sample, the number of independent coordinates in these equations can be reduced by one, from six to five. Further simplification of Boltzmann equations is achieved by applying the first-order angular moments expansion to the charge densities. These simplified equations can then be treated numerically in an efficient way.

We will solve these simplified Boltzmann equations, and show how efficiently they can follow the dynamics of electrons. We will consider two study cases of a very different dynamics: (i) the case where Coulomb electrostatic interactions between charged particles are included, (ii) the case where we neglect electrostatic forces in the system. The results obtained will be discussed in detail. Afterwards, a short summary will be given. Finally, we will list our conclusions.

# 2 Boltzmann equation

Statistical description of a classical system can be made in terms of its density function, $\rho(\mathbf{r}, \mathbf{v}, t)$ [13–15], where $\rho(\mathbf{r}, \mathbf{v}, t)$ is defined such that $\rho(\mathbf{r}, \mathbf{v}, t)d^3rd^3v$ is the number of particles at time t positioned between $\mathbf{r}$ and $\mathbf{r} + \mathbf{dr}$ which have velocities in the range $(\mathbf{v}, \mathbf{v} + \mathbf{dv})$.

Consider a classical gas of identical particles in which an external force, $\mathbf{F}$, acts, and assume initially that no collisions take place between the gas particles. Within the time, $dt$, the velocity $\mathbf{v}$ of any particle will change to $\mathbf{v} + \mathbf{F}dt/m$ and its position, $\mathbf{r}$, will change to $\mathbf{r} + \mathbf{v}dt$. As the system is assumed to conserve number of particles, the number of particles at the time, $t$, $\rho(\mathbf{r}, \mathbf{v}, t)d^3rd^3v$, will be equal to the number of particles at the time, $t + dt$, $\rho(\mathbf{r} + \mathbf{v}dt, \mathbf{v} + \mathbf{F}dt/m, t + dt)d^3rd^3v$:

$$\rho(\mathbf{r} + \mathbf{v}dt, \mathbf{v} + \mathbf{F}dt/m, t + dt)d^3rd^3v - \rho(\mathbf{r}, \mathbf{v}, t)d^3rd^3v = 0 \qquad (1)$$

If, however, collisions do occur between the particles, or if there are some other processes of particle creation or annihilation, there will be a difference between the number of particles at the time, $t$, and the number of particles at the time, $t + dt$. This change is then described by a collision (source) operator, $\Omega(\rho, \mathbf{r}, \mathbf{v}, t)$.

In the limit, $dt \to 0$, Eq. (1) transforms to the following general equation describing the evolution of the density function:

$$\partial_t \rho + \mathbf{v}\partial_{\mathbf{r}}\rho + \mathbf{F}\partial_{\mathbf{v}}\rho/m = \Omega(\rho, \mathbf{r}, \mathbf{v}, t). \qquad (2)$$

This is the famous Boltzmann equation. Its most important feature is the ability to describe non-equilibrium processes. Boltzmann equations are used for describing transport processes in many different physical contexts, ranging from the simulations of the hot electron transport in semiconductors, simulations of plasma kinetics [12, 16–19] to the evolution of protoneutron stars [20] and to the modelling of the core collapse in supernovas [21].

Boltzmann equations conserve the number of particles, the total energy and the momenta of a closed system. The equilibrium solution of Boltzmann equations yields the Maxwell-Boltzmann distribution of velocities [22]. This distribution is also an asymptotic solution of the non-equilibrium Boltzmann equations at long time scales, $t \to \infty$.

We will now formulate the specific Boltzmann equations describing the transport of electrons, atoms and ions inside a sample irradiated with FEL photons. In this case it is enough to consider two gases: the gas of light electrons of masses, $m$ and charges, $-e$, and the gas of heavy atoms/ions of masses, $M$, and charges, $ie$. Photons need not to be considered as an independent gas component, as they only enter the equations as a flux term in the photoionization source term. The gases of electrons and atoms/ions are described by the density functions: $\rho^{(e)}(\mathbf{r}, \mathbf{v}, t)$, $\rho^{(i)}(\mathbf{r}, \mathbf{v}, t)$, where $i$ denotes the ion charge $i = 0, 1, \ldots, N_J$, and $N_J$ is an arbitrary number, describing the maximal ion charge in the system. The general coupled Boltzmann equations for these gases are:

$$\partial_t \rho^{(e)}(\mathbf{r}, \mathbf{v}, t) + \mathbf{v} \cdot \partial_{\mathbf{r}} \rho^{(e)}(\mathbf{r}, \mathbf{v}, t) + \frac{e}{m} \left( \mathbf{E}(\mathbf{r}, t) + \mathbf{v} \times \mathbf{B}(\mathbf{r}, t) \right) \cdot \partial_{\mathbf{v}} \rho^{(e)}(\mathbf{r}, \mathbf{v}, t) = \Omega^{(e)}(\rho^{(e)}, \rho^{(i)}, \mathbf{r}, \mathbf{v}, t), \quad (3)$$

for electrons, and

$$\partial_t \rho^{(i)}(\mathbf{r}, \mathbf{v}, t) + \mathbf{v} \cdot \partial_{\mathbf{r}} \rho^{(i)}(\mathbf{r}, \mathbf{v}, t) - \frac{ie}{M} \left( \mathbf{E}(\mathbf{r}, t) + \mathbf{v} \times \mathbf{B}(\mathbf{r}, t) \right) \cdot \partial_{\mathbf{v}} \rho^{(i)}(\mathbf{r}, \mathbf{v}, t) = \Omega^{(i)}(\rho^{(e)}, \rho^{(i)}, \mathbf{r}, \mathbf{v}, t), \quad (4)$$

for atoms/ions, where the force $\mathbf{F}$ is the electromagnetic force, $\mathbf{F}(\mathbf{r}, \mathbf{v}, t) = q(\mathbf{E}(\mathbf{r}, t) + \mathbf{v} \times \mathbf{B}(\mathbf{r}, t))$, acting on electrons and ions positioned between $\mathbf{r}$ and $\mathbf{r} + \mathbf{dr}$, which have velocities in the range $(\mathbf{v}, \mathbf{v} + \mathbf{dv})$. The electric field, $\mathbf{E}$, and magnetic field, $\mathbf{B}$, have two components. The first component describes the interaction of charges with external radiation. The second component describes internal electromagnetic interaction between electrons and ions. This component is a non-local function of electron and ion densities.

Collision (source) terms, $\Omega^{(e,i)}$, describe the change of the electron/ion densities of velocities $(\mathbf{v}, \mathbf{v} + \mathbf{dv})$ measured at the positions $(\mathbf{r}, \mathbf{r} + \mathbf{dr})$ with time. This change is due to: (i) the creation of the secondary electrons and highly charged ions via photo- and collisional ionizations of atoms and ions, (ii) elastic and inelastic collisions of electrons and ions. If the collision terms are neglected, Boltzmann equations, Eqs. (3), (4), reduce to the Vlasov equation [12, 16] describing the evolution of a collisionless plasma.

Initial configuration of Eqs. (3), (4) is given by a smooth atomic density function, $\rho^{(0)}(\mathbf{r}, \mathbf{v}, 0)$, which represents the sample at $t = 0$.

## 3 Boltzmann equation for an irradiated atomic cluster

First experimental studies on the interaction of the intense VUV photon beams with matter were performed for clusters of xenon atoms irradiated with the VUV photons [8]. New experiments with clusters exposed to the FEL radiation at higher photon energies are planned in the next future. The existing and the future experimental data give a unique opportunity for testing the applicability of different theoretical models.

Below we formulate the assumptions of the primary transport model dedicated for studying the dissipative dynamics and the radiation damage in xenon clusters at the VUV photon energies. We will fix the physical parameters as they were set in the first experiment with the VUV photons [8].

The production terms, $\Omega^{(e,i)}$, in our model will then include only basic predominant interactions, i.e.:

(i) **Single photoionizations of atoms**. A single VUV photon of energy, $E_\gamma = 12.7$ eV, may excite electrons only from the $5p_{3/2}$ shell of xenon atoms of the binding energy, $E_i = 12.1$ eV. Here the photon energy was set as in the VUV FEL experiment [8]. We neglect possible multistep photo- and multiphoton ionizations within this primary model.

(ii) **Elastic and inelastic collisions of electrons and atoms/ions**. We assume that an inelastic collision always releases a secondary electron. We neglect inelastic collisions of electron and atoms/ions which lead only to an excitation of an atom/ion. These processes contribute to the multistep collisional ionization which is not included within this primary model.

(iii) **Inverse bremsstrahlung photoabsorption in the presence of atoms or ions**. In our model, as the primary kinetic energy of a photoelectron released by a VUV photon is small, $E \sim 0.6$ eV, comparing to the first ionization energy, $E_i = 12.1$ eV, an efficient process of energy pumping is necessary in order to initiate any collisional ionizations by electrons. Inverse bremsstrahlung process is among the possible processes [6]. Here, the bremsstrahlung is treated in a classical way as an energy gain by an electron inside the sample which is due to the interaction of this electron with the electric field of the laser. This electron-laser field interaction is treated within the dipole approximation.

(iv) **Electromagnetic interactions of electrons and ions within the sample**. They are expressed in the form of the non-local potentials. We assume for simplicity that both electrons and ions are point-like charged particles, and neglect the effects of the atom/ion internal structure and of its finite size. Including these effects would require a modified atomic potential which we do not include into this primary model.

Finally, we note that within this primary model we also neglect the recoil energies and the recoil momenta of the atoms/ions gained during their interactions with photons or electrons. Electrons are assumed to scatter isotropically on atoms/ions. These are the first order approximations which can be made: (i) in case of photoionizations due to low energy of the incoming photons, and (ii) in case of collisional interactions due to large difference of electron and ion masses and low impact energies of electrons. Within this approximation the movement of ions will be stimulated by the Coulomb repulsion only, and will start at the final stages of the explosion. The additional pressure on ions due to the recoil momenta is neglected. Recoil effects, and also short-range electron-electron interaction may be conveniently treated by the means of the Fokker-Planck equation. As other relevant processes which were neglected within this primary model, e.g. three-body recombination, charge enhanced ionization [9] or effects of electron screening on the atomic potentials [6, 23], these processes will be treated in forthcoming papers.

## 3.1 General Boltzmann equations for electrons and ions in an irradiated sample

Before writing the equations, we will introduce the following notation. The integrated densities, $\overline{\rho}^{(e,i)}(\mathbf{r},t)$ are defined as,

$$\overline{\rho}^{(e,i)}(\mathbf{r},t) = \int d^3v\, \rho^{(e,i)}(\mathbf{r},\mathbf{v},t). \tag{5}$$

Velocity $v_E = \sqrt{2(E_\gamma - E_i)/m}$ is the magnitude of the velocity of the photoelectrons. Coefficients $\sigma_\gamma^{i\to i+1}$ denote the total photoionization cross sections for a single ionization of an ion of charge, $i = 0, 1, \ldots$. Coefficients $\sigma_{ic}^{i\to i+1}$ denote the total collisional cross sections for a single ionization of an ion of charge, $i = 0, 1, \ldots$ by an electron. Coefficients $\sigma_{ec}^{i\to i}$ denote the elastic cross sections for electron-ion collisions. For xenon elastic electron-atom scattering cross sections and ionization electron-atom/ion cross sections were measured experimentally [24–28]. Within this primary model the cross section for the elastic electron-ion scattering was approximated by the cross section for the elastic electron-atom scattering, and isotropic scattering of electrons on ions was assumed.

A compact notation for doubly differential cross sections is used, $\frac{d\sigma^{i\to j}(\boldsymbol{v_e};\boldsymbol{v'_e}(\boldsymbol{v_s}))}{d\mathbf{v}}$. Velocity $\mathbf{v}_e$ denotes the velocity of the incoming electron, $\mathbf{v}'_e$ is the velocity of this electron after the collision, $\mathbf{v}_s$ is the velocity of the secondary electron.

Coefficient $j(E_\gamma)$ describes the photon flux, and $\Omega$ is a spherical angle.

Starting from the Eqs. (3), (4), we derived the following equations for electron and ion densities in the irradiated sample:

$$
\begin{aligned}
& \frac{\partial \rho^{(e)}(\mathbf{r},\mathbf{v},t)}{\partial t} + \mathbf{v}\frac{\partial \rho^{(e)}(\mathbf{r},\mathbf{v},t)}{\partial \mathbf{r}} - \\
- \quad & \frac{\partial \rho^{(e)}(\mathbf{r},\mathbf{v},t)}{\partial \mathbf{v}} \cdot \left( \frac{e^2}{4\pi\epsilon_0 m} \int d^3r' \, \frac{\mathbf{r}-\mathbf{r}'}{\mid \mathbf{r}-\mathbf{r}' \mid^3} \left\{ \sum_{i=0}^{N_J} i \cdot \overline{\rho}^{(i)}(\mathbf{r}',t) - \overline{\rho}^{(e)}(\mathbf{r}',t) \right\} + \frac{e}{m} E(t)\, \boldsymbol{\epsilon} \right) = \\
= \quad & \sum_{i=0}^{N_J} \overline{\rho}^{(i)}(\mathbf{r},t)\, j(E_\gamma) \frac{d\sigma_\gamma^{i\to i+1}(E_\gamma;\mathbf{v})}{d\Omega_{\mathbf{v}}} \frac{\delta(v-v_E)}{v^2} + \\
+ \quad & \sum_{i=0}^{N_J} \overline{\rho}^{(i)}(\mathbf{r},t) \left\{ \int d^3v_e \, v_e \, \rho^{(e)}(\mathbf{r},\mathbf{v_e},t) \frac{d\sigma_{ec}^{i\to i}(\mathbf{v_e};\mathbf{v})}{d\Omega_{ev}} \frac{\delta(v-v_e)}{v^2} - v\rho^{(e)}(\mathbf{r},\mathbf{v},t)\, \sigma_{ec}^{i\to i}(\mathbf{v}) \right\} + \\
+ \quad & \sum_{i=0}^{N_J} \overline{\rho}^{(i)}(\mathbf{r},t) \left\{ \int d^3v_e \, v_e \, \rho^{(e)}(\mathbf{r},\mathbf{v_e},t) \left( \frac{d\sigma_{ic}^{i\to i+1}(\mathbf{v_e};\mathbf{v_e'}=\mathbf{v})}{d\mathbf{v}} + \frac{d\sigma_{ic}^{i\to i+1}(\mathbf{v_e};\mathbf{v_s}=\mathbf{v})}{d\mathbf{v}} \right) - \right. \\
& \qquad \left. -v\rho^{(e)}(\mathbf{r},\mathbf{v},t)\sigma_{ic}^{i\to i+1}(\mathbf{v}) \right\} \qquad (6)
\end{aligned}
$$

for electrons and:

$$
\begin{aligned}
& \frac{\partial \rho^{(i)}(\mathbf{r},\mathbf{v},t)}{\partial t} + \mathbf{v}\frac{\partial \rho^{(i)}(\mathbf{r},\mathbf{v},t)}{\partial \mathbf{r}} + \\
+ \quad & \frac{\partial \rho^{(i)}(\mathbf{r},\mathbf{v},t)}{\partial \mathbf{v}} \cdot \left( \frac{e^2}{4\pi\epsilon_0 M} \int d^3r' \, \frac{\mathbf{r}-\mathbf{r}'}{\mid \mathbf{r}-\mathbf{r}' \mid^3} \left\{ \sum_{j=0}^{N_J} (ij) \cdot \overline{\rho}^{(j)}(\mathbf{r}',t) - i\overline{\rho}^{(e)}(\mathbf{r}',t) \right\} + \frac{ie}{M} E(t)\, \boldsymbol{\epsilon} \right) = \\
= \quad & j(E_\gamma)\sigma_\gamma^{i-1\to i}(E_\gamma)\rho^{(i-1)}(\mathbf{r},\mathbf{v},t) - j(E_\gamma)\sigma_\gamma^{i\to i+1}(E_\gamma)\rho^{(i)}(\mathbf{r},\mathbf{v},t) + \\
+ \quad & \left\{ \int d^3v_e \, \sigma_{ic}^{i-1\to i}(\mathbf{v_e}) \, v_e \, \rho^{(e)}(\mathbf{r},\mathbf{v_e},t) \right\} \rho^{(i-1)}(\mathbf{r},\mathbf{v},t) - \\
- \quad & \left\{ \int d^3v_e \, \sigma_{ic}^{i\to i+1}(\mathbf{v_e}) \, v_e \, \rho^{(e)}(\mathbf{r},\mathbf{v_e},t) \right\} \rho^{(i)}(\mathbf{r},\mathbf{v},t) \qquad (7)
\end{aligned}
$$

for ions, where we treated the interaction with the external electric field within the dipole approximation (inverse bremsstrahlung),

$$\mathbf{E}(\mathbf{r},t) \cong E(t)\boldsymbol{\epsilon}, \qquad (8)$$

and neglected the subleading contribution coming from the interaction of charges with magnetic field. Vector, $\boldsymbol{\epsilon}$, is the polarization vector of the electric field.

Writing these equations we took into account only binary collisions between participating particles. The assumption of binary collisions is not valid for very dense systems and for systems with the presence of long-range forces, where many body effects become important [11]. Within this primary model we neglected short-range three and higher many body interactions ocurring due to the high density of particles in the sample. Many body effects due to the presence of Coulomb forces were treated correctly within the approximation that recoil energies and recoil momenta of atoms and ions could be neglected.

Equations (6), (7) then describe the evolution of a sample irradiated with VUV FEL photons within our primary model. The structure of these equations is general, and other interactions or improvements can be easily implemented into these equations. These completed equations would then describe a more advanced model of the sample dynamics. Eqs. (6), (7) can be also adapted for describing the dynamics of an irradiated sample at other photon energies.

## 3.2 Solving the Boltzmann equations

Equations (6), (7) are complicated integro-differential equations in six-dimensional phase space. They can be treated only numerically.

For a spherically symmetric cluster the number of dimensions can be reduced by one, from six to five. A significant simplification of the Boltzmann equations (6), (7) can be achieved by expanding the electron and ion densities in terms of their angular moments. This method was successfully applied for the description of the evolution of the protoneutron stars [20] and plasmas [12, 16]. An assumption has then to be made that the isotropic components of the electron and ion densities are predominant. Here we mean isotropy in phase space, and not only in space, i.e. at each spatial point of an isotropic spatial distribution the velocity distribution has also to be isotropic. Such approximate isotropy occurs in systems where there is a strong collisional dissipation of particle energies, and the phase space component of the collective transport is small. This is certainly the case for low energy electrons inside an ionic/atomic cluster, as they then frequently collide (with short range forces) with ions and atoms inside this cluster.

The validity of the angular moments method may be also verified a posteriori, i.e. one may compare the magnitude of the isotropic and angular components of the charge densities after solving the Boltzmann equations. If the isotropic components of the electron and ion densities obtained with these equations were much larger that the other components of the electron densities, the angular moments approximation worked correctly, and the solution of the equation obtained within this approximation was a good estimate of the full solution.

# 4 Test of Boltzmann method on study case: simplified electron dynamics inside atomic cluster

## 4.1 Simplifying assumptions

In order to test the applicability and efficiency of the Boltzmann method for describing the dynamics of electrons in FEL irradiated samples, we have applied it to a test case of a simplified electron dynamics.

Our initial configuration was given by a smooth atomic density function, representing a spherically symmetric cluster consisting of $909$ neutral xenon atoms. This cluster was irradiated with the VUV FEL photons of energies, $E_\gamma = 12.7$ eV. For simplicity we have also assumed that the photon pulse had a constant intensity, and it was switched on instantaneously at $t = 0$ fs. The pulse intensity corresponded to an estimate of the maximal FEL pulse intensity observed in the experiment [8], $I = 10^{14}$ W/cm$^2$. The pulse length was, $\Delta t = 50$ fs.

We have applied the following simplifying assumptions to the electron and ion/atom dynamics. First, we have expanded the electron density using the angular moment expansion:

$$\rho^{(e)}(\mathbf{r}, \mathbf{v}, t) \cong \frac{1}{2\pi} \left( \rho_0^{(e)}(r, v, t) + cos(\theta_{rv}) \cdot \rho_1^{(e)}(r, v, t) \right), \tag{9}$$

and within this diffusion approximation kept only: (i) its zeroth order (isotropic) component, $\rho_0^{(e)}(r, v, t)$, which corresponds to the number of electrons inside the volume element $dV = d^3r\, d^3v$, and (ii) its first order transport component, $\rho_1^{(e)}(r, v, t)$, which contributes to the particle flux through the borders of the phase space element. [3] Radius, $r$, is the distance from the centre of the cluster ($r =\mid \mathbf{r} \mid$), and $v$ denotes the magnitude of the electron velocity ($v =\mid \mathbf{v} \mid$). The function $cos(\theta_{rv})$ denotes the cosine of the relative angle between vectors $\mathbf{v}$ and $\mathbf{r}$. The isotropic component of the electron density, $\rho_0^{(e)}(r, v, t)$, has to be a positive number, as it describes the number of electrons in an infinitesimal volume element, $dV = d^3r\, d^3v$. The transport component of the density, $\rho_1^{(e)}(r, v, t)$, can be a positive or a negative number. Positive values of $\rho_1^{(e)}$ indicate that there is a collective transport in phase space outwards the cluster, the negative ones indicate that there is a collective transport inwards the cluster.

[3] In case of central forces the total particle flux in real space through the sphere of radius, $r$, is: $S(r) = \frac{8\pi}{3} \int_0^\infty dv\, v^3\, r^2\, \rho_1^{(e)}(r, v, t)$, and the total flux in velocity space through the sphere of radius, $v$, is: $S(v) = \frac{8\pi}{3} \int_0^\infty dr\, a(r)\, v^2\, r^2\, \rho_1^{(e)}(r, v, t)$, where $a(r)$ is the radial acceleration.

The approximation (9) is valid if the densities, $\rho_0^{(e)}(r,v,t)$ and $\rho_1^{(e)}(r,v,t)$ fulfill the following condition:

$$\rho_0^{(e)}(r,v,t) >> \mid \rho_1^{(e)}(r,v,t) \mid \tag{10}$$

We also note that there is no coupling of the electron density to the laser field within this simplified model due to the dipole approximation (8) and diffusion approximation (9).

Second, as we are only interested in following the electron dynamics within this simplified model, we further assume that the positions of ions are fixed and their velocities remain equal to zero during the evolution:

$$\rho^{(i)}(\mathbf{r},\mathbf{v},t) \cong \frac{1}{2\pi}\rho_0^{(i)}(r,v,t)\cdot\frac{\delta(v)}{v^2}. \tag{11}$$

In the true physical case this assumption is valid only during the first stages of the exposure. The approximation (11) implies that $\overline{\rho}^{(i)}(\mathbf{r},t) = 2\rho_0^{(i)}(r,v=0,t)$.

Within the diffusion approximation (9) Boltzmann equations for the electron density, Eqs. (6), reduce to:

$$\begin{aligned}
&\frac{\partial \rho_0^{(e)}(r,v,t)}{\partial t} + \frac{v}{3r^2}\cdot\frac{\partial(r^2\rho_1^{(e)}(r,v,t))}{\partial r} - \frac{A(r,\rho_0^{(i)},\rho_0^{(e)})}{3v^2}\cdot\frac{\partial(v^2\rho_1^{(e)}(r,v,t))}{\partial v} = \\
&= \sum_{i=0}^{N_J}\rho_0^{(i)}(r,0,t)\left\{ j(E_\gamma)\sigma_\gamma^i(E_\gamma)\frac{\delta(v-v_E)}{v^2} + 2\int_0^\infty dv_e\, v_e^3\,\rho_0^{(e)}(r,v_e,t)\,d\sigma_{tot}^i(v_e;v)\right. \\
&\qquad \left. -2v\rho_0^{(e)}(r,v,t)\sigma_{tot}^i(v)\right\}
\end{aligned}$$

$$\begin{aligned}
&\frac{\partial \rho_1^{(e)}(r,v,t)}{\partial t} + v\cdot\frac{\partial \rho_0^{(e)}(r,v,t)}{\partial r} - A(r,\rho_0^{(i)},\rho_0^{(e)})\cdot\frac{\partial \rho_0^{(e)}(r,v,t)}{\partial v} = \\
&= -\sum_{i=0}^{N_J} 2\rho_0^{(i)}(r,0,t)\,v\rho_1^{(e)}(r,v,t)\sigma_{tot}^i(v),
\end{aligned} \tag{12}$$

where:

$$A(r,\rho_0^{(i)},\rho_0^{(e)}) = \frac{e^2}{4\pi\epsilon_0 m}\cdot\frac{8\pi}{r^2}\cdot\int_0^r dr'\,r'^2\left\{\sum_{i=0}^{N_J} i\cdot\rho_0^{(i)}(r',0,t) - \int_0^\infty dv\,v^2\,\rho_0^{(e)}(r',v,t)\right\}, \tag{13}$$

$$d\sigma_{tot}^i(v_e;v) = \sigma_{ec}^{i\to i}(v_e)\frac{\delta(v-v_e)}{v^2} + \frac{2}{v^2}\frac{d\sigma_{ic}^{i\to i+1}(v_e;v)}{dv}, \tag{14}$$

$$\sigma_{tot}^i(v) = \sigma_{ec}^{i\to i}(v) + \sigma_{ic}^{i\to i+1}(v), \tag{15}$$

and $\sigma_\gamma^i(E_\gamma) \equiv \sigma_\gamma^{i\to i+1}(E_\gamma)$. The Coulomb electrostatic force in Eq. (12) has been expanded using the multipole expansion with the accuracy consistent with the accuracy of the diffusion approximation (9). Electron-ion and electron-atom collisions were assumed to be isotropic which is a good approximation at low impact energies of electrons. For the simulation purpose the delta-like photoionization velocity distribution, $\frac{\delta(v-v_E)}{v^2}$ in Eq. (12), has to be approximated with a gaussian profile of a non-zero width and a mean value at the photoelectron velocity, $v_E$. The normalization constant of the gaussian profile is chosen in order to maintain the correct number of the photoelectrons released.

As there is no transport of ions within our approximation, the zeroth order component of the ion density, $\rho_0^{(i)}(r,v,t)$, is sufficient to represent the full ion density, $\rho^{(i)}(\mathbf{r},\mathbf{v},t)$ (Eq. (11)). Therefore equation (7) reduces to an ordinary rate equation, describing the change of the number of ions due to the photo- and collisional ionizations:

$$\begin{aligned}\frac{\partial\rho_0^{(i)}(r,0,t)}{\partial t} &= \rho_0^{(i)}(r,0,t)\left\{j(E_\gamma)\sigma_\gamma^{i-1}(E_\gamma)+2\int_0^\infty dv_e\, v_e^3\,\rho_0^{(e)}(r,v_e,t)\,\sigma_{ic}^{i-1}(v_e)\right\}-\\ &- \rho_0^{(i)}(r,0,t)\left\{j(E_\gamma)\sigma_\gamma^{i}(E_\gamma)+2\int_0^\infty dv_e\, v_e^3\,\rho_0^{(e)}(r,v_e,t)\,\sigma_{ic}^{i}(v_e)\right\}, \qquad (16)\end{aligned}$$

where $i=1,2,\ldots,N_J$. For atomic densities this equation simplifies to:

$$\frac{\partial\rho_0^{(0)}(r,0,t)}{\partial t} = -\rho_0^{(0)}(r,0,t)\left\{j(E_\gamma)\sigma_\gamma^{0}(E_\gamma)+2\int_0^\infty dv_e\, v_e^3\,\rho_0^{(e)}(r,v_e,t)\,\sigma_{ic}^{0}(v_e)\right\} \qquad (17)$$

## 4.2 Results

We have prepared a dedicated algorithm for solving the Boltzmann equations including the relevant numerical methods [29–31]. Integrals and partial derivatives in these equations were evaluated using the pseudospectral method [30]. This algorithm has been extensively tested, e.g. the interactions terms were included step-by-step into the code, the energy and the particle number conservation were monitored during the evolution. The accuracy of the time integration has been checked with two independent time-integration methods: (i) of the fifth, and (ii) of the third order. Obtaining the predictions for a single case took several hours on the AlphaStation XP1000.

We then solved Eqs. (12), (16) and (17) numerically in two study cases of very different electron dynamics: (i) with Coulomb electrostatic interactions between charges included, (ii) with Coulomb interactions not included. We followed the evolution of the cluster up to $50$ fs of the exposure. Photon energy was $E_\gamma = 12.7$ eV. The simulation box had the size: ($0 < r < 120$ A)$\times$ ($0 < v < 30$ A/fs), and it was divided into 40$\times$70 grid points respectively. The simulation box corresponded to a sphere in real space of radius, $r = 120$ A, and a sphere of radius, $v = 30$ A, in velocity space. This box was surrounded by an absorbing wall. Figs. 1-11 show the results.

The quantities obtained after solving Boltzmann equations were: the three-dimensional electron density functions, $\rho_0^{(e)}(r,v,t)$, $\rho_1^{(e)}(r,v,t)$, and the integrated two-dimensional ion/atom distributions, $\overline{\rho}^{(i)}(\mathbf{r}, t)$, recorded at different times, $t = 0, \ldots, 50$ fs. Figs. 1,2 show an example of the isotropic and the transport component of the electron density in phase space obtained with Boltzmann equations at time, $t = 2$ fs. Plotted are the functions: $\tilde{\rho}_j^{(e)}(r, v, t) = r^2\, v^2\, \rho_j^{(e)}(r, v, t)$, where $j = 0, 1$. As expected, the isotropic component of the electron density function, $\tilde{\rho}_0^{(e)}(r, v, t)$, is a positively defined function (Fig. 1). This function is localized in phase space (see the contour plot). In contrast, the transport component of the electron density, $\tilde{\rho}_1^{(e)}(r, v, t)$, may take both positive and negative values. In the contour plot of Fig. 2, the upper part of the contour at $v = 4 - 8$ A/fs with a peak at negative values of $\tilde{\rho}_1^{(e)}(r, v, t)$ indicates the inward transport. The lower part of the contour plot at $v = 1 - 3$ A/fs with a peak at positive values of $\tilde{\rho}_1^{(e)}(r, v, t)$ indicates the outward transport.

These three-dimensional plots are not easy to analyze. More transparent information on the evolution of the electron cloud can be obtained from plots of the integrated isotropic and transport density functions, $n$, defined as,

$$\begin{aligned} n_j(v, t) &\equiv \int \rho_j^{(e)}(r, v, t)\; r^2\, dr, \\ n_j(r, t) &\equiv \int \rho_j^{(e)}(r, v, t)\; v^2\, dv. \end{aligned} \tag{18}$$

The integrated isotropic component, $n_0(v, t)$ and $n_0(r, t)$ are related to the full integrated densities as, $\overline{\rho}^{(e)}(\mathbf{r}, t) = 2\, n_0(r, t)$, and, $\overline{\rho}^{(e)}(\mathbf{v}, t) = 2\, n_0(v, t)$, within the diffusion approximation (9). For ions we have: $n_0^{(i)}(r, t) \equiv \rho_0^{(i)}(r, 0, t)$. The total number of electrons (ions), $N^{(e,i)}(t)$, can then be obtained after performing the integration of the isotropic component of the density function over $d^3r\, d^3v$:

$$8\pi \cdot \int \rho_0^{(e,i)}(r, v, t)\; r^2\, v^2\, dr\, dv = N^{(e,i)}(t). \tag{19}$$

Figs. 3 and 7 show the atomic, single ion and double ion density functions in two cases: (i) with Coulomb forces included, and (ii) with Coulomb forces not included. In both cases almost all photoelectrons have been released within the first femtosecond of the exposure. This result is consistent with the photoionization rate estimated at this photon energy, the assumed pulse intensity and the pulse shape. Atomic density significantly decreased within the first femtosecond of the exposure. In contrast, the single ion density at $t = 1$ fs followed the shape of the initial atomic distribution at $t = 0$ fs. There were no double or highly charged ions observed in the cluster, as the electrons released were not energetic enough for further collisional ionizations. Also, photons were assumed to induce single photoionizations of the neutral atoms only.

The dynamics of electrons was very different in these two study cases. In the first case Coulomb forces were preventing most of the electrons from leaving the ionic cluster. Some of the electrons were, however, able to escape. The largest flows of energy and particles have been observed within $5 - 15$ fs of the exposure (Fig. 6d). Within this time weak collective oscillations of the electron cloud around the ion cloud were observed (not shown). After this time electrons thermalized, and their anisotropic transport component became negligible. Fig. 4 shows the rapid progress of the thermalization process for both the isotropic and the transport components of the integrated electron density, $n_0(v,t)$ and $n_1(v,t)$. The initial free electron density was equal to zero. After the first femtosecond of the exposure the shape of the isotropic electron density, $n_0(v,t)$, followed the gaussian profile of the photoelectron velocity distribution, and it broadened with time. Full thermalization was achieved at times $\geq 10$ fs within this test model. We have fitted Maxwell-Boltzmann distribution to the results on the integrated density function, $n_0(v,t)$, obtained at $t = 20$ fs (Fig. 4): $n_0(v,t) = a \cdot exp(-mv^2/(2k_BT))$. The electron temperature was estimated to, $k_BT = 0.68 - 0.77$ eV, which corresponded to the average energy, $\langle E \rangle = 3k_BT/2 = 1.02 - 1.15$ eV. This value agreed well with the average energy estimated with the global parameters at $t = 20$ fs (Fig. 6a,c), $\langle E \rangle \equiv E_{kinet}/N_{el} = 1.07$ eV.

Here we recall that the energy transferred to the system by a single photon of energy, $E_\gamma = 12.7$ eV, was $E_{ph-el} = 1.1$ eV instead of $0.6$ eV, as we had to approximate the delta-like photoionization velocity distribution, $\frac{\delta(v-v_E)}{v^2}$ in Eq. (12) with a gaussian profile of a non-zero width and a mean value at the photoelectron velocity, $v_E$. The normalization constant of the gaussian profile was chosen in order to maintain the correct number of the photoelectrons released. The energy of a single photoelectron obtained with the integral of this modified velocity distribution was then equal to $1.1$ eV.

Fig. 4b shows the time evolution of the transport component of the electron density, $n_1(v,t)$, corresponding to the weak plasma oscillations. There is a strong increase of the outward electron transport (in velocity space) within 2 fs of the exposure. Energetic electrons can then leave the simulation box (Fig. 6d). However, at some time point, $t \sim 5$ fs, slower electrons travelling outward are stopped and attracted back by ions. The inward transport start then to dominate. After the thermalization of the electrons is achieved, the collective transport (in velocity space) reduces significantly, and the transport component of the electron cloud becomes small.

Spatial evolution of the electron cloud (Fig. 5), described by the integrated densities, $n_0(r,t)$ and $n_1(r,t)$, is less dynamic than the evolution of the velocity densities, $n_0(v,t)$ and $n_1(v,t)$. A rapid increase of $n_0(r,t)$ is observed only within the first femtosecond of the exposure. After this time, the shape of the isotropic component of the integrated electron density does not change much with time. Weak plasma oscillations are visible, if $r^2\, n_0(r,t)$ is plotted (not shown). The magnitude of the spatial transport component of the density function, $n_1(r,t)$, is much smaller than its isotropic component, $n_0(r,t)$, during the evolution.

During the first femtoseconds of the exposure there is a weak spatial transport outward, and the position of the maximum of the transport component propagates towards lower values of $r$ at increasing times. This corresponds to a plasma wave propagating inside the cluster. These weak oscillations occur until there are damped at about $20$ fs of the exposure. This damping is due to the thermalization of electrons.

In Fig. 6 we plot also global parameters of the sample as functions of time: the total, kinetic and potential energy of the system, the particle number, and the flows of energy and the particle numbers recorded at a fixed distance of 10 grid points from the external borders of the simulation box. These flows give a valuable qualitative information about the escape rate of the electrons at different stages of the evolution. This information is helpful for estimating the correct size of the simulation box. If the box size would be too small, some electrons of a total negative energy could leave the box during the evolution. This would lead to a strong increase of the potential energy within the system and induce an incorrect electron dynamics.

Total kinetic energy of the cluster and the number of electrons and ions increased rapidly within the $2$ fs of the exposure. Within and after this time we observed a strong outward flow of electrons. The fastest electrons were able to leave the simulation box. Potential energy slowly increased with time, as more electrons escaped from the simulation box. It was, however, small if compared to the total energy. Strong oscillations of the energy flow during the escape phase were reflected by the shape of the potential energy curve.

In the totally unrealistic case (ii), where the electrostatic interactions between particles have not been included, electrons were observed to drift slowly towards the external borders of the simulation box (Figs. 8, 9). Their velocity distribution remained unchanged within the first few femtoseconds of the exposure. After the electrons left the ionized cluster ($r > 25$ A), their dissipative elastic collisions with ions were no longer possible. The collective movement of the electrons lead to the fast breakdown of the diffusion approximation. This breakdown occurred already within $10$ fs of the exposure. The isotropicity condition (10) was then no longer fulfilled, and this showed up in the integrated electron density distributions, $n_0(v,t)$ and $n_0(r,t)$. They decreased with time, as expected, because more and more electrons were leaving the simulation box. However, at times, $t > 10$ fs, these densities took incorrect negative values at small values of $v$ or $r$ (Figs. 8, 9a). Also, some global parameters were affected by the breakdown of the diffusion approximation. After $10$ fs both the total (kinetic) energy of the system and the electron number strongly decreased with time down to some unphysical (negative) values at about $25$ fs of the exposure (Fig. 10).

In order to check that our results obtained in case (i) were not a numerical artifact, we have also tested the response of the system in case of a small change of the photon energy, $E_\gamma = 14$ eV. This change of energy increased twice the energy pumped into the system by a single photon, $E_{ph-el} = 2.4$ eV. Here we only show the global parameters recorded as functions of time (Fig. 11). Their time characteristics is similar to that obtained in case of the lower photon energy, $E_\gamma = 12.7$ eV. However, as the photoelectrons here were more energetic, the potential energy, energy flows and also particle flows oscillated more violently with time within the first tens of femtoseconds of the exposure. Also, more electrons were able to leave the simulation box. As in the previous case, the potential energy of the system slowly increased after the escape phase at times $\geq 15$ fs. All these results were consistent with the results obtained at the photon energy, $E_\gamma = 12.7$ eV.

# Summary and conclusions

We have formulated general Boltzmann equations describing the evolution of a non-uniform sample irradiated with the FEL photons. We have solved these equations numerically with a dedicated algorithm in two study cases of simplified electron dynamics inside a spherically symmetric xenon cluster. This cluster was irradiated with a high intensity pulse of VUV FEL radiation. The predictions obtained in case when internal Coulomb forces were included gave a comprehensive description of the evolution of the electron cloud during its non-equilibrium (before thermalization) and equilibrium stages (after thermalization). At photon energies, $E_\gamma = 12.7$ eV, thermalization of electrons was observed within our system after $10$ fs of the exposure. This thermalization was an effect of the non-local long-range Coulomb forces, and not of the interparticle collisions, as the energy transfers in the non-ionizing electron collisions were not allowed within our model.

During the exposure almost all photoelectrons were confined inside the ion cluster by the Coulomb internal field. Only a few of photoelectrons were energetic enough to escape. Therefore the electrostatic energy of the sample was too low to accelerate remaining electrons and initiate further collisional ionizations. These results confirm that other effective mechanisms of energy pumping are necessary for the creation of highly charged ions in the sample irradiated with VUV photons, e.g. inverse bremsstrahlung or multistep ionization.

The model developed here needs significant improvements of its physical assumptions in order to be applied to a realistic case. The mechanisms of energy pumping (inverse bremstrahlung process and multi- or multistep ionizations) should be included, and their influence on the overall ionization dynamics carefully examined. Important mechanisms of thermalization: recoil effects and short-range electron-electron interactions, need to be included into the extended model. Possible influence of recombination and of other many body processes on the sample dynamics requires careful analysis. Including all these interactions into Boltzmann equations will require going beyond the diffusion approximation.

However, we do not expect that including further interactions into the equations will lead to more numerical complicacies. As the main nonlinearity and stability problems have been successfully treated in the primary algorithm, and this algorithm correctly followed the dynamics of the sample in our study case, we expect that it can easily be extended for a more advanced model.

We believe that the method proposed here offers a unique possibility of studying the complex dynamics of large samples, irradiated with the FEL pulses. Whereas in real experiments the sample is exposed to several processes contributing simultaneously to the radiation damage, the Boltzmann simulation tool enables one to include specific interactions only. In this way the influence of different ionization mechanisms on the overall dynamics of the sample can conveniently be tested.

To sum up, Boltzmann approach is a first principle model which can follow non-equilibrium processes in phase space. Single particle densities evolved with Boltzmann equations include the full information on particle positions and velocites, and not only on their collective components. Average observables obtained with the Boltzman solver are not biased with statistical errors. However, the information on the three and higher order correlations is not included within Boltzmann equations. Including the effects of many body correlations into these equations is generally not possible, only in a few cases and under simplifying assumptions, e.g. by applying the Fokker-Planck equation in case of long-range Coulomb forces. The other serious disadvantage of the Boltzmann approach is its numerical complexity which requires an application of advanced numerical methods.

Computational costs within Boltzmann approach do not scale with the number of atoms within a sample, as in the MC method. Therefore, a Boltzmann solver is usually more efficient for larger samples than a Monte Carlo code. However, computational costs in Boltzmann approach depend on the number of grid points in phase space. Apparently, this limits the size and shape of samples which can be studied with Boltzmann equations. Applying the Boltzmann solver for describing the evolution of a large sample of a very irregular shape would be extremely time-consuming.

To sum up, we have demonstrated that the Boltzmann equations are a powerful tool to follow the radiation damage of a non-uniform sample irradiated with the FEL photons. We believe that these equations may soon become a standard tool for investigating the complex dynamics of irradiated samples.

## Acknowledgements

Beata Ziaja is grateful to T. Chmaj, J. Feldhaus, S. Hau-Riege, L. Motyka, R. A. London and B. Sonntag for illuminating discussions. She thanks G. Ingelman for providing an access to the computing units of the high energy physics group at the Uppsala University. This research was supported in part by the Polish Committee for Scientific Research with grant No. 2 P03B 02724. Beata Ziaja is a fellow of the Alexander von Humboldt Foundation.

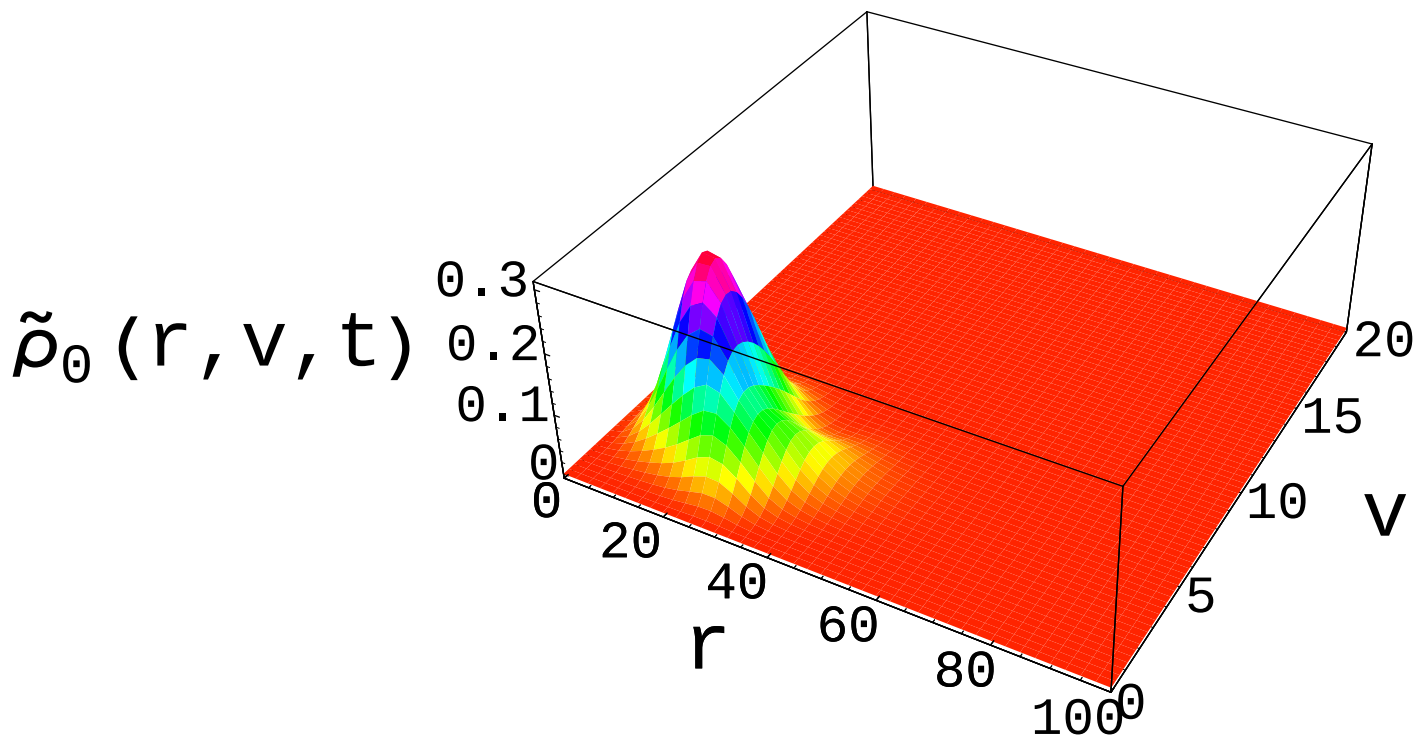

a)

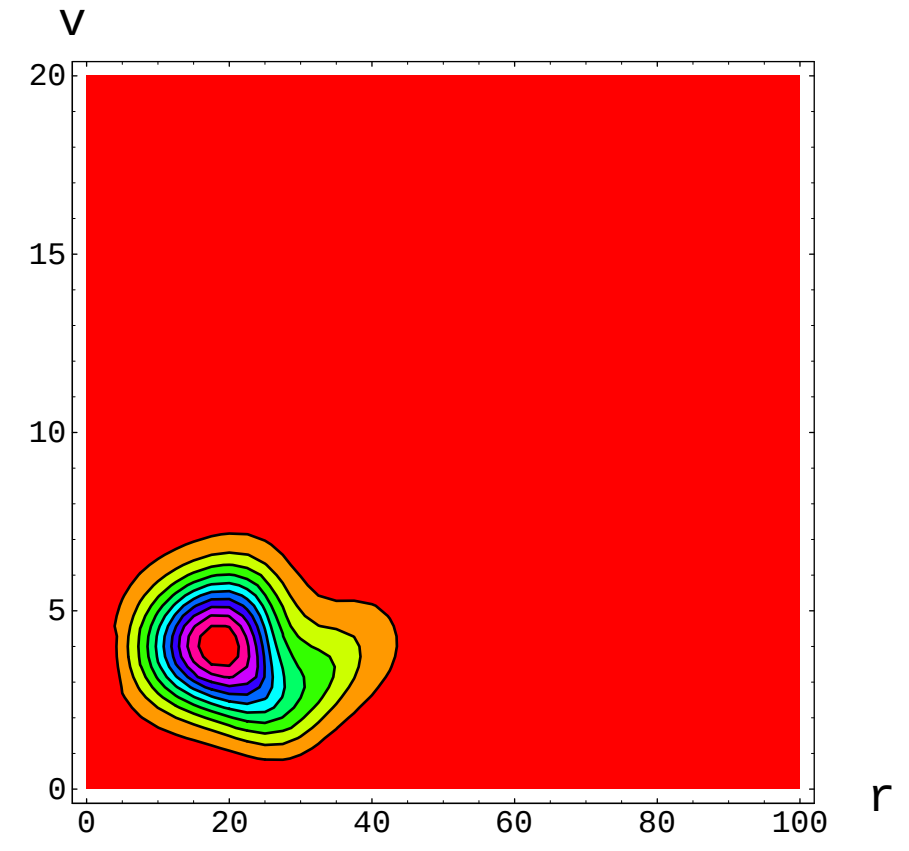

b)

Figure 1: Isotropic component of the electron density in phase-space, $\tilde{\rho}_0(r, v, t) = r^2\, v^2\, \rho_0(r, v, t)$, recorded at time, $t = 2$ fs: a) three-dimensional view and b) contour plot. This density was obtained in case of the irradiation with the VUV FEL photons of energies, $E_\gamma = 12.7$ eV. Coulomb interactions between charged particles were included. Initial density of free electrons at $t = 0$ fs was equal to $0$. The ranges of axes correspond to the size of the simulation box.

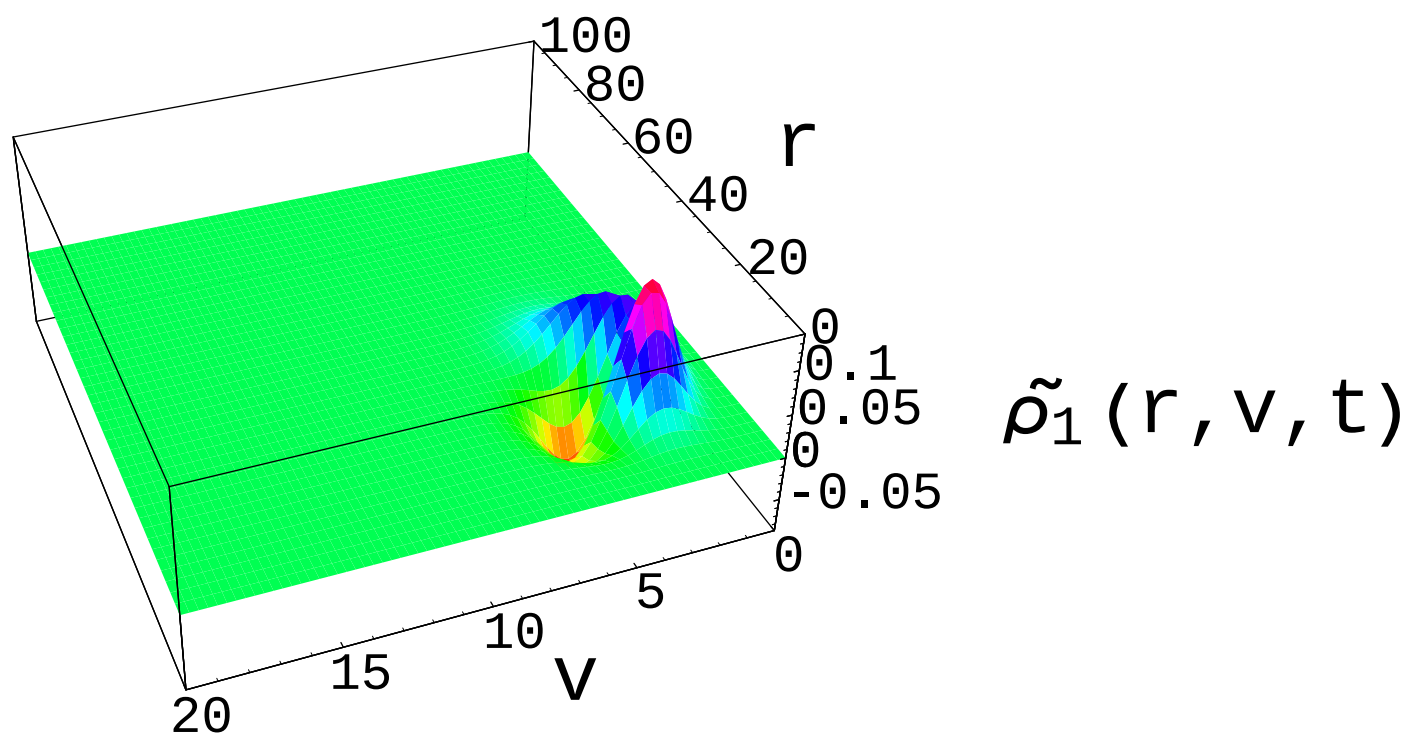

a)

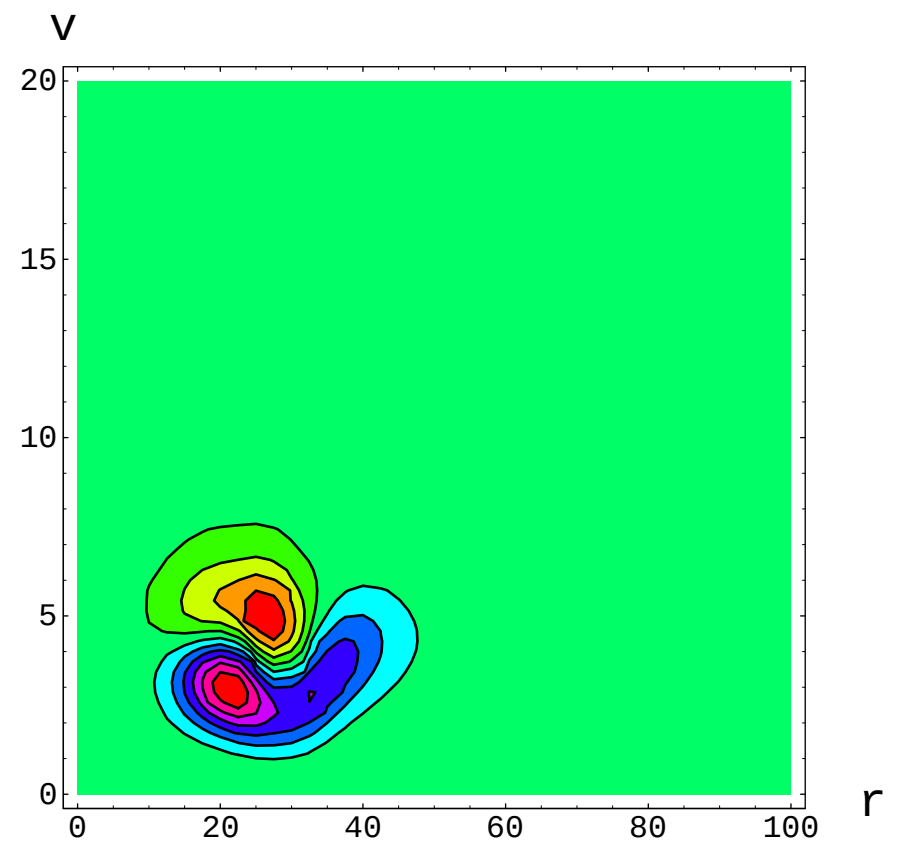

b)

Figure 2: Transport component of the electron density in phase-space, $\tilde{\rho}_1(r, v, t) = r^2\, v^2\, \rho_1(r, v, t)$, recorded at time, $t = 2$ fs: a) three-dimensional view and b) contour plot. This density was obtained in case of the irradiation with the VUV FEL photons of energies, $E_\gamma = 12.7$ eV. Coulomb interactions between charged particles were included. Initial density of free electrons at $t = 0$ fs was equal to $0$. The ranges of axes correspond to the size of the simulation box.

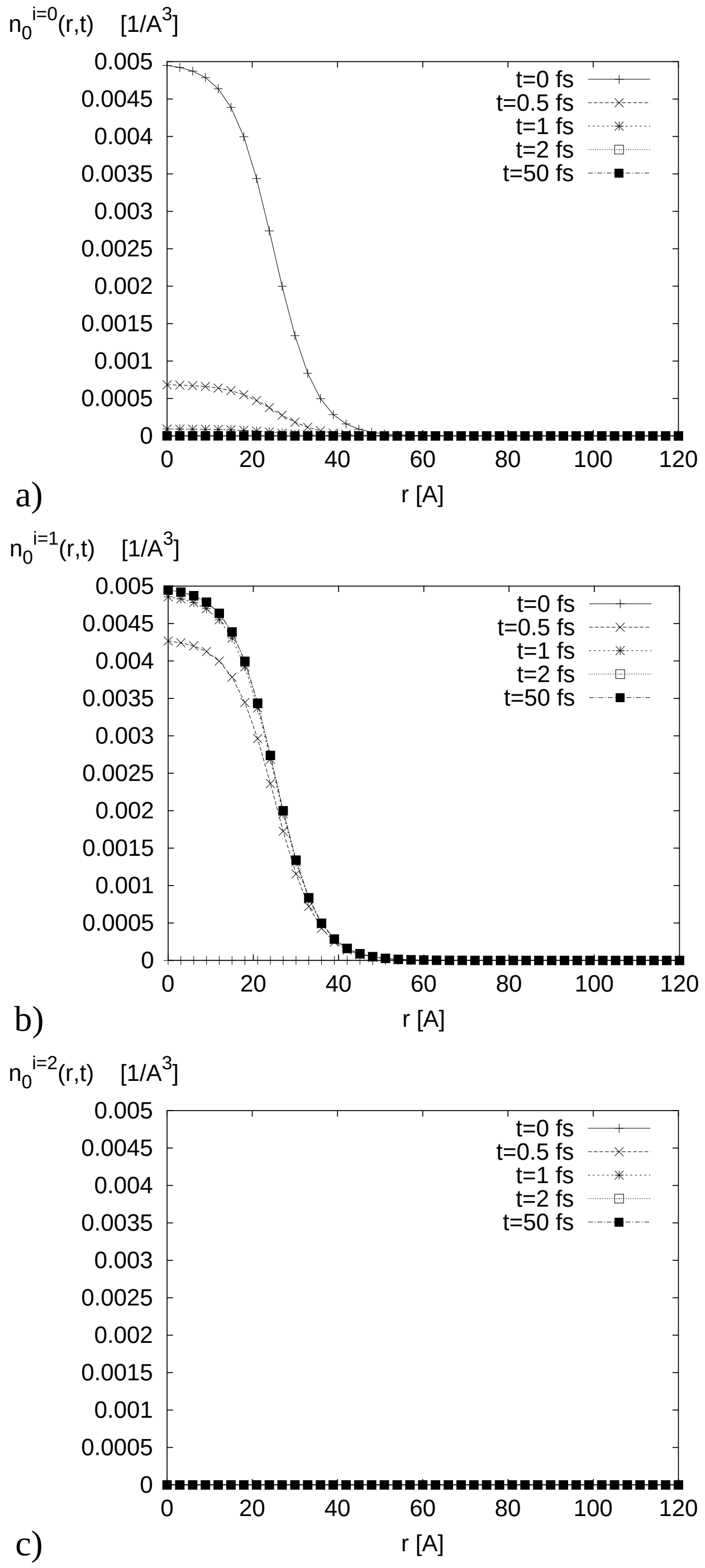

Figure 3: Integrated atom and ion densities, $n_0^{i=0,1,2}(r,t)$: a) atomic density, b) single ion density, c) double ion density, recorded at times, $t = 0, \ldots, 50$ fs. These densities were obtained in case of the irradiation with the VUV FEL photons of energies, $E_\gamma = 12.7$ eV. Coulomb interactions between charged particles were included.

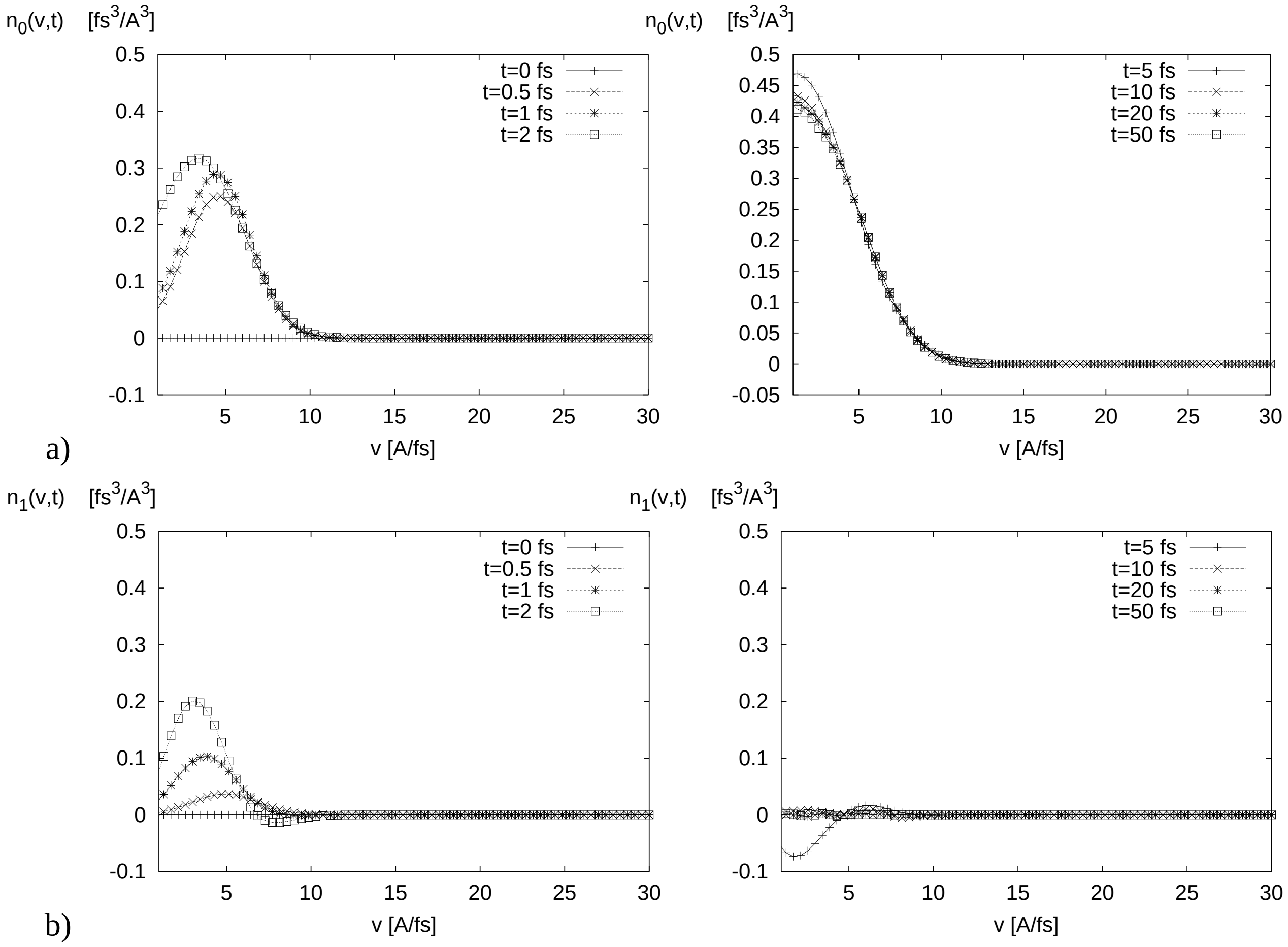

Figure 4: Integrated electron density: a) isotropic component, $n_0(v,t)$, b) transport component, $n_1(v,t)$, recorded at different times, $t = 0, \ldots, 50$ fs. These densities were obtained in case of the irradiation with the VUV FEL photons of energies, $E_\gamma = 12.7$ eV. Coulomb interactions between charged particles were included.

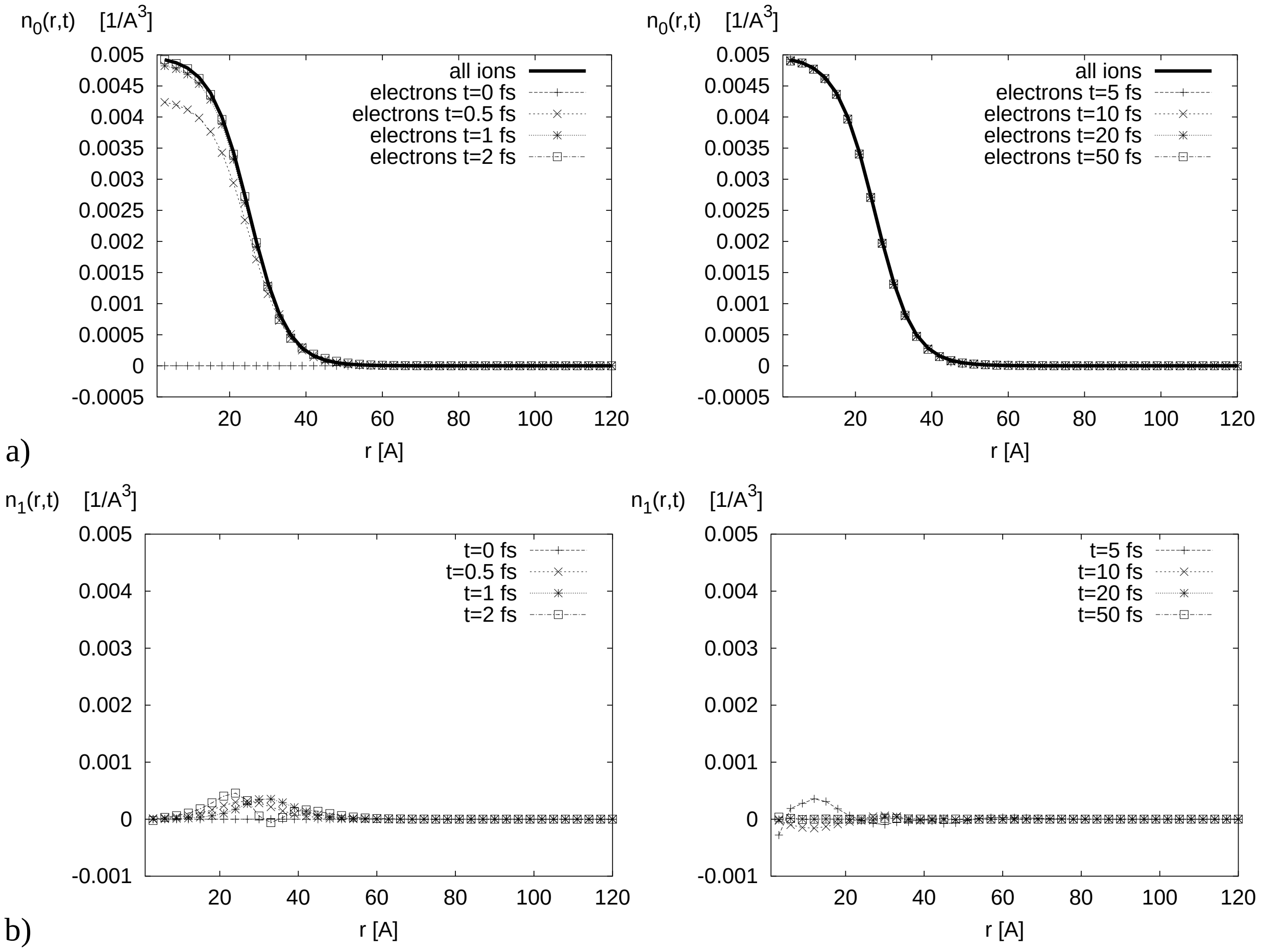

Figure 5: Integrated electron density: a) isotropic component, $n_0(r,t)$, b) transport component, $n_1(r,t)$, recorded at different times, $t = 0, \ldots, 50$ fs. These densities were obtained in case of the irradiation with the VUV FEL photons of energies, $E_\gamma = 12.7$ eV. Coulomb interactions between charged particles were included.

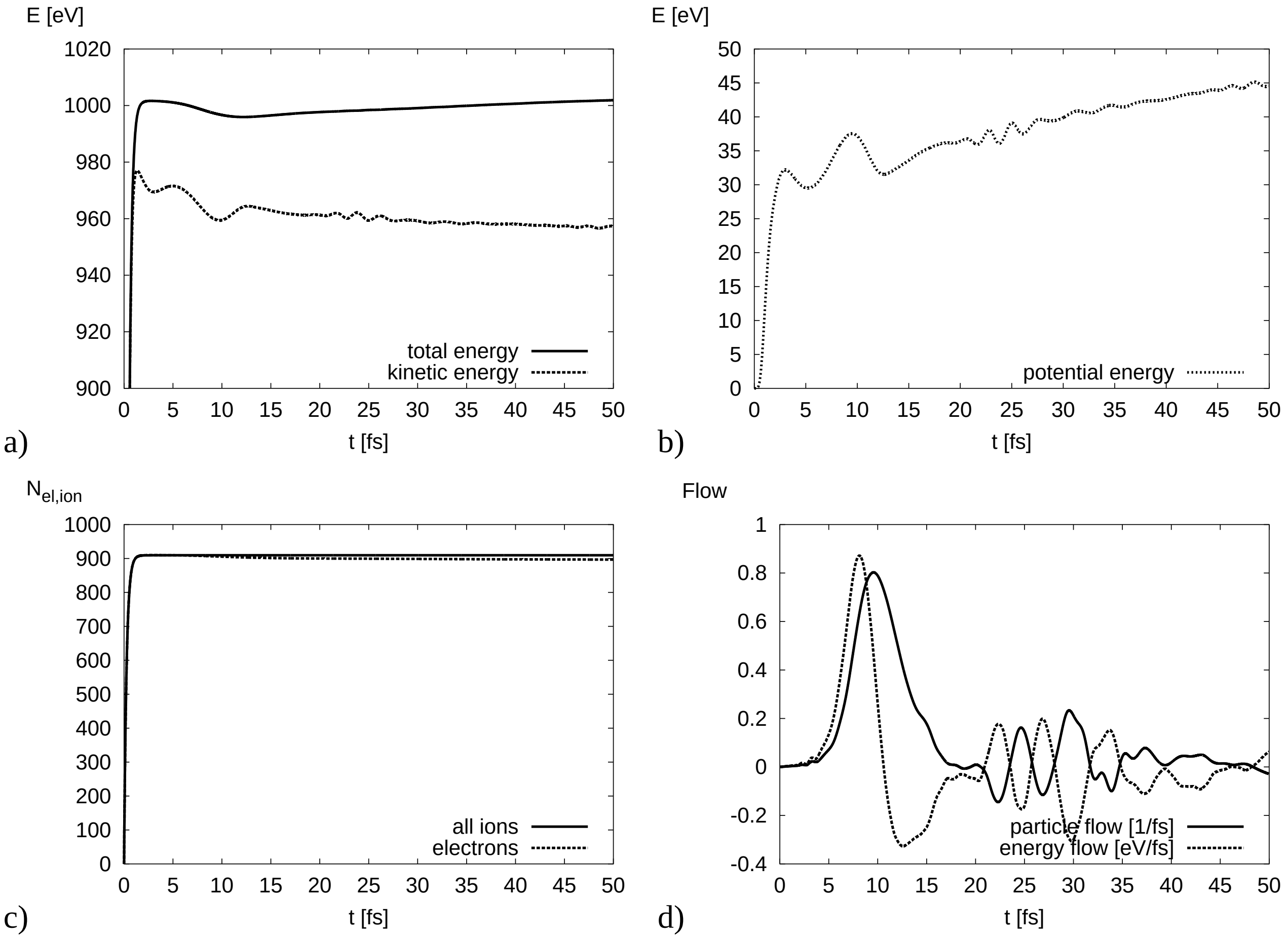

Figure 6: Global parameters of the irradiated cluster as functions of time: a) total energy, b) potential energy, c) number of electrons and singly charged ions in the cluster, d) flows of energy and particles measured at the distance of 10 grid points from the external borders of the simulation box. These parameters were obtained in case of the irradiation with the VUV FEL photons of energies, $E_\gamma = 12.7$ eV. Coulomb interactions between charged particles were included.

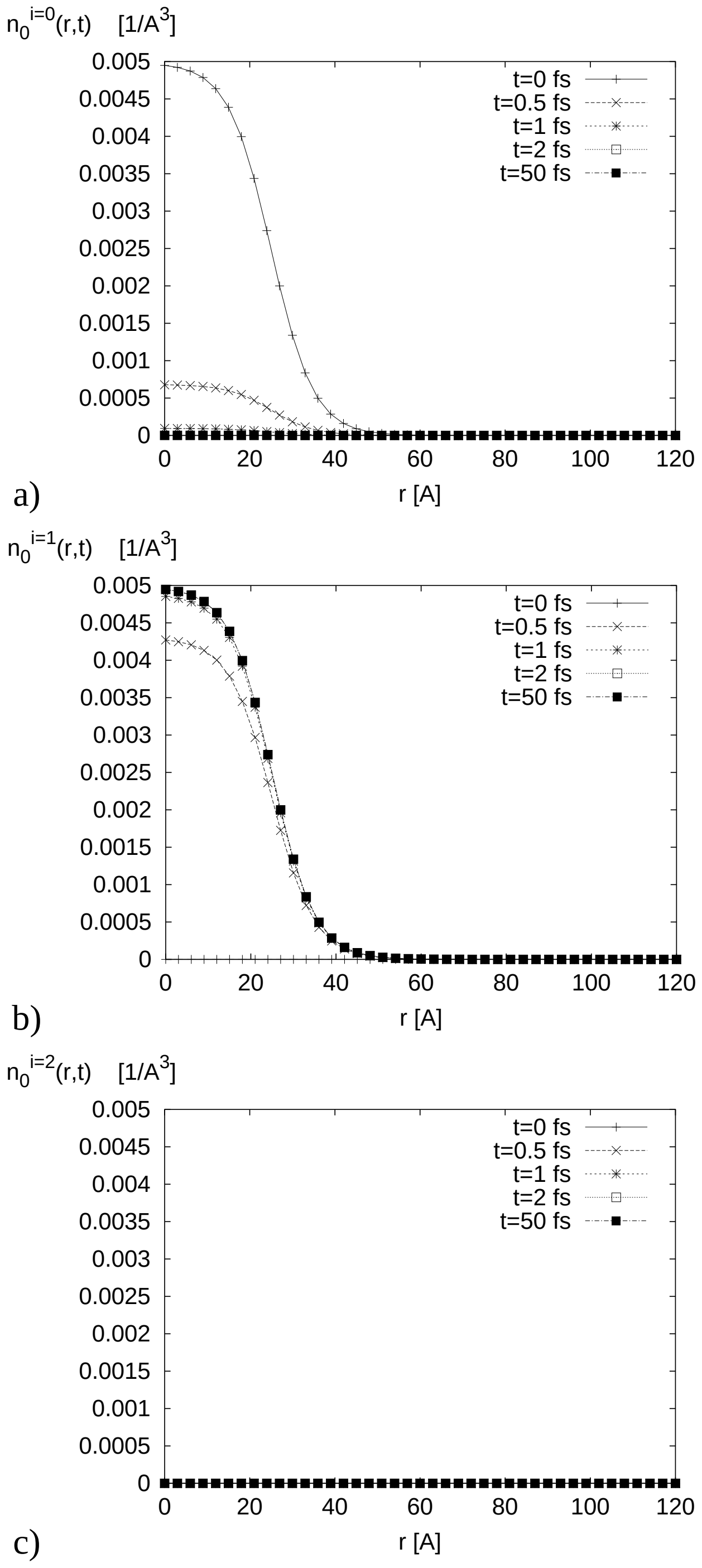

Figure 7: Integrated atom and ion densities, $n_0^{i=0,1,2}(r,t)$: a) atomic density, b) single ion density, c) double ion density, recorded at times, $t = 0, \ldots, 50$ fs. These densities were obtained in case of the irradiation with the VUV FEL photons of energies, $E_\gamma = 12.7$ eV. Coulomb interactions between charged particles were **not** included.

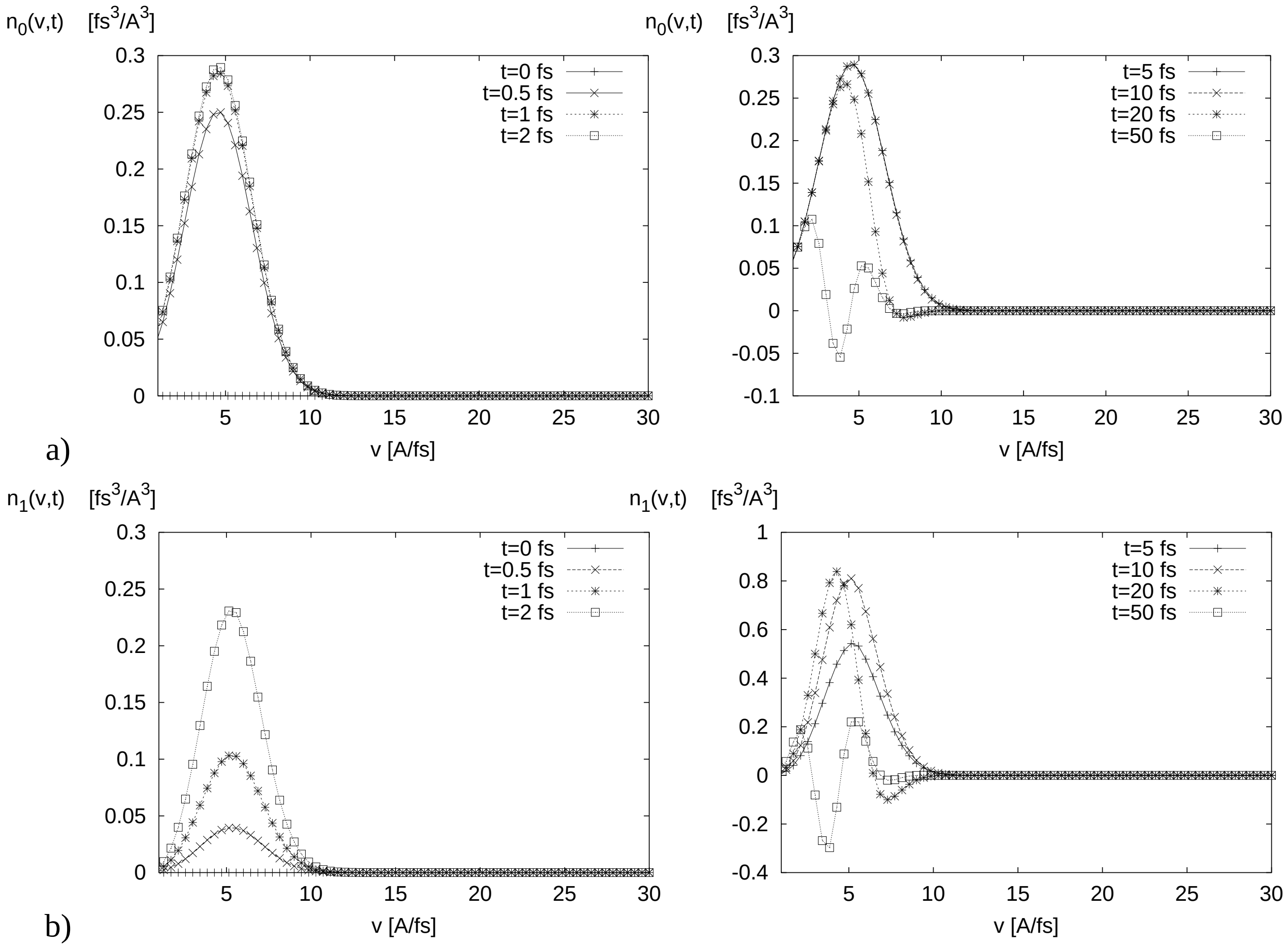

Figure 8: Integrated electron density: a) isotropic component, $n_0(v,t)$, b) transport component, $n_1(v,t)$, recorded at different times, $t = 0, \ldots, 50$ fs. These densities were obtained in case of the irradiation with the VUV FEL photons of energies, $E_\gamma = 12.7$ eV. Coulomb interactions between charged particles were **not** included.

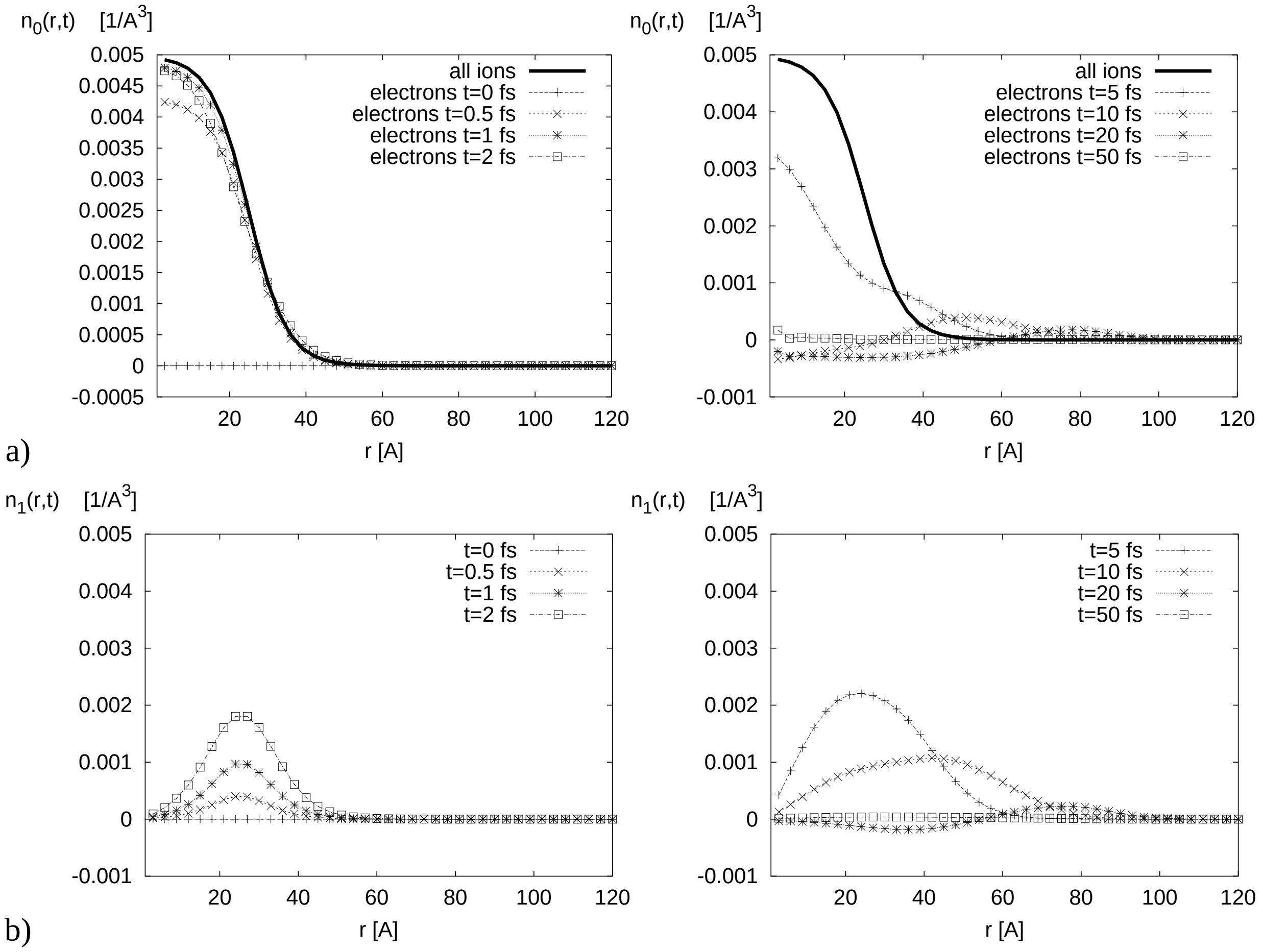

Figure 9: Integrated electron density: a) isotropic component, $n_0(r,t)$, b) transport component, $n_1(r,t)$, recorded at different times, $t = 0, \ldots, 50$ fs. These densities were obtained in case of the irradiation with the VUV FEL photons of energies, $E_\gamma = 12.7$ eV. Coulomb interactions between charged particles were **not** included.

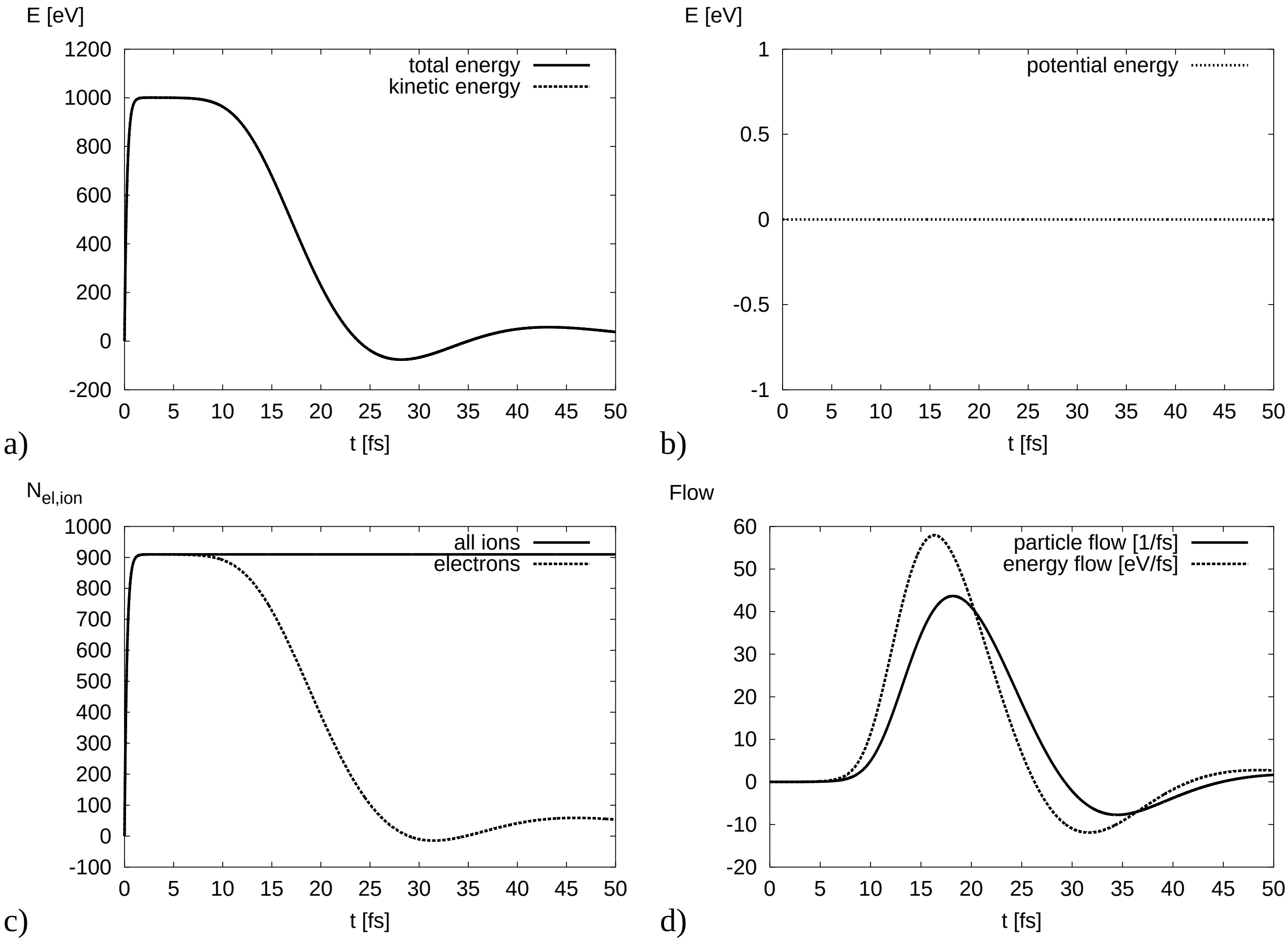

Figure 10: Global parameters of the irradiated cluster as functions of time: a) total energy, b) potential energy, c) number of electrons and singly charged ions in the cluster, d) flows of energy and particles measured at the distance of 10 grid points from the external borders of the simulation box. These parameters were obtained in case of the irradiation with the VUV FEL photons of energies, $E_\gamma = 12.7$ eV. Coulomb interactions between charged particles were **not** included.

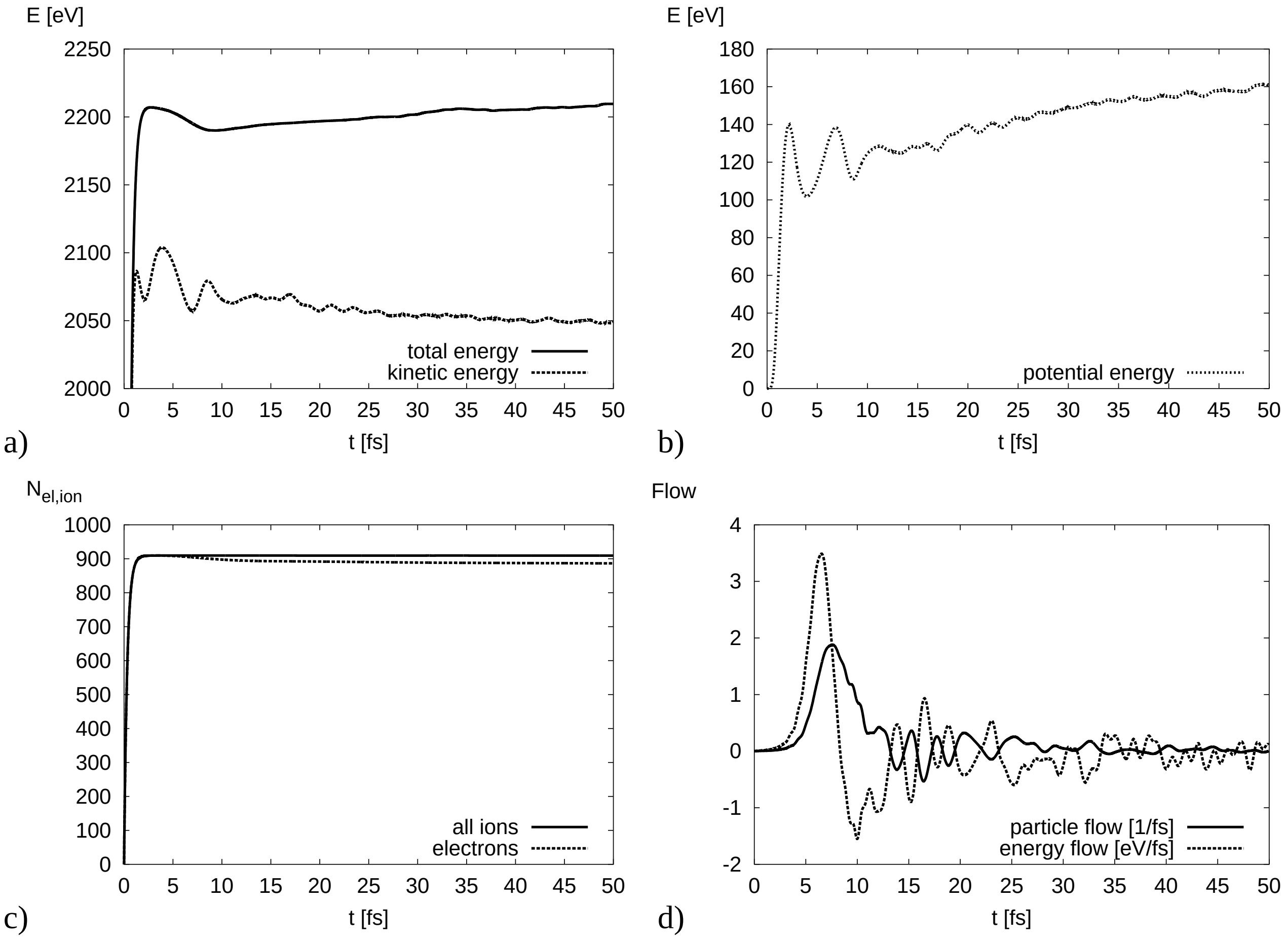

Figure 11: Global parameters of the irradiated cluster as functions of time: a) total energy, b) potential energy, c) number of electrons and singly charged ions in the cluster, d) flows of energy and particles measured at the distance of 10 grid points from the external borders of the simulation box. These parameters were obtained in case of the irradiation with the VUV FEL photons of energies, $E_\gamma = 14$ eV. Coulomb interactions between charged particles were included.